\documentclass[journal,twoside,twocolumn]{IEEEtran}

\usepackage{float}       
\usepackage{cite}
\usepackage{graphicx}
\usepackage{algorithm}
\usepackage{algorithmic}
\usepackage{booktabs}
\usepackage{array}
\usepackage{xcolor}
\usepackage{hyperref}
\usepackage{mathtools}
\usepackage{url}
\usepackage[none]{hyphenat}
\usepackage{multirow}
\usepackage{amssymb}

\DeclareMathOperator{\CRLB}{CRLB}

\begin{document}
\title{LEO Signals-of-Opportunity for Navigation in Urban Environments: A GLRT-Assisted Off-Grid SBL Approach}

\author{Maedeh~Sotodenia, Mohammad Hossein Kahaei, Alireza Nezamalhosseini%
\thanks{School of Electrical Engineering, Iran University of Science and Technology.
(e-mail:maedeh\_sotodenia@elec.iust.ac.ir;
kahaei@iust.ac.ir;	nezam@iust.ac.ir).
}}

%\begin{document}
%\title{\fontsize{16}{19}\selectfont
%LEO Signals of Opportunity for Navigation in Urban Environments:
%A GLRT-Assisted Off-Grid SBL Approach}
%\author{Maedeh~Sotodenia}
%\affil{*****}

%\author{Mohammad Hossein Kahaei}
%\affil{****}

%\author{Alireza Nezamalhosseini}
%\affil{*****}

%\authoraddress{School of Electrical Engineering, Iran University of Science and Technology, Tehran, Iran
%(e-mail: maedeh\_sotodenia@elec.iust.ac.ir;
%kahaei@iust.ac.ir; nezam@iust.ac.ir).}

%\maketitle

\maketitle
\begin{abstract}
The vulnerability and unavailability of global navigation satellite system (GNSS) signals in dense urban environments have motivated the use of low earth orbit (LEO) signals of opportunity (SoOP) for positioning, navigation, and timing (PNT). However, multipath and non-ideal noise can significantly degrade delay--Doppler estimation and, consequently, navigation accuracy. This paper proposes a hybrid generalized likelihood ratio test (GLRT) and off-grid sparse Bayesian learning (SBL) framework for LEO-SoOP navigation under multipath and non-Gaussian, temporally correlated noise. First, a GLRT-based framework is developed, demonstrating reliable detection under ideal Gaussian noise but degraded estimation under heavy-tailed or colored noise and unresolved multipath. Next, an off-grid SBL framework is introduced for high-resolution separation and estimation of line-of-sight (LOS) delay--Doppler parameters. The estimated LOS parameters are subsequently tracked and incorporated into an extended Kalman filter (EKF) to obtain the navigation solution. Simulation results demonstrate that, in urban environments, the proposed method reduces delay RMSE by up to 77\% and position RMSE by 88.5\% compared with GLRT-based estimation. These results demonstrate the potential of the proposed framework for robust and accurate LEO-SoOP navigation.

\end{abstract}

\section{Introduction}
\label{sec:introduction}
Positioning, navigation, and timing (PNT) services are essential to modern applications, including transportation, logistics, precision agriculture, and military operations. global navigation satellite systems (GNSS) have long provided these services and have become critical infrastructure. However, GNSS is vulnerable to jamming, spoofing, and signal blockage or loss of visibility, particularly in dense urban and indoor environments~\cite{zidan2020gnss,rados2024gnss}. These vulnerabilities, together with recent GNSS outages, have highlighted the need for independent, robust, resilient, and cost-effective complementary or alternative PNT solutions~\cite{stock2025survey}.
\begin{figure}
    \centering
    \includegraphics[width=0.9\linewidth]{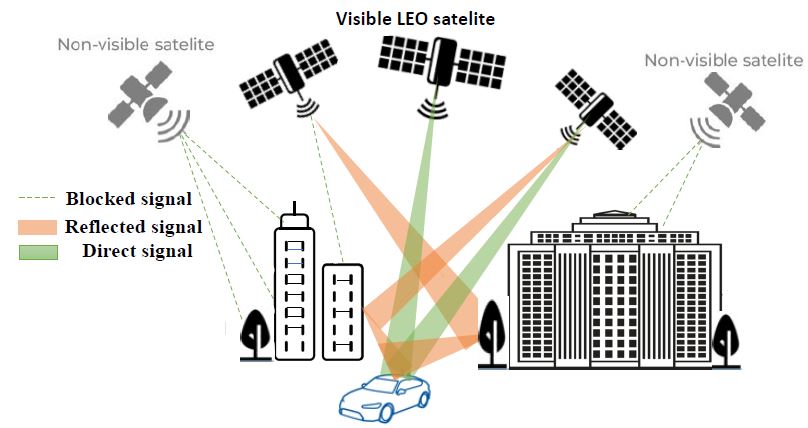}

    \label{fig:enter-label}
    
    \caption{Urban propagation environment for LEO-SoOP navigation.}
    
\end{figure}

 An effective approach to resilient PNT is to exploit signals of opportunity (SoOP), including terrestrial radio and cellular networks and non-terrestrial LEO satellite signals. LEO satellites are particularly attractive because their lower orbits provide stronger received signals and larger Doppler shifts than GNSS MEO satellites, benefiting positioning and velocity estimation~\cite{kassas2024adastra}. The increasing number of visible LEO satellites also provides greater geometric and frequency diversity and improved signal availability~\cite{stock2025survey}.

Despite the advantages of LEO-SoOP signals, several challenges limit their effective use for PNT. First, LEO satellite ephemerides are often poorly known. Although two-line element (TLE) data can be propagated using the simplified general perturbations 4 (SGP4) model~\cite{vallado2008sgp4}, orbit prediction errors can cause significant position and velocity uncertainties, degrading navigation observables. Second, unlike GNSS, LEO satellites may not have highly stable atomic clocks. Third, the signal structures of commercial LEO satellites are often not publicly available or fully known, making signal acquisition, tracking, and extraction of reliable navigation observables more challenging~\cite{kassas2024adastra,khairallah2024ephemeris,shahcheraghi2025acq,neinavaie2021blind}. Fourth, in urban environments, LEO-SoOP signals are severely affected by multipath propagation from buildings and other obstacles. These multipath components distort the delay--Doppler correlation structure, complicate LOS detection and acquisition, and introduce biased delay and Doppler estimates that can degrade positioning performance~\cite{abdallah2021multipath,kassas2024truth}.

Several blind and cognitive techniques enable navigation with non-cooperative LEO signals by extracting navigation observables without full knowledge of the transmitted waveform. Their feasibility has been demonstrated with signals from Orbcomm, Starlink, and OneWeb using blind Doppler estimation and opportunistic receiver architectures~\cite{kassas2024adastra,shahcheraghi2025acq,neinavaie2021blind}.
\begin{table*}[!t]
\centering
\caption{Comparison of representative studies with the proposed framework
for LEO SoOP-based navigation.}
\label{tab:literature_comparison}
\renewcommand{\arraystretch}{0.7}
\setlength{\tabcolsep}{4pt}
\resizebox{\textwidth}{!}{%
\begin{tabular}{l|c|c|c|c|c|c}
\toprule
\textbf{Work}
& \textbf{LEO--SoOP}
& \textbf{Urban Multipath}
& \textbf{Off-grid SBL}
& \textbf{Robust Noise}
& \textbf{LOS/NLOS Separation}
& \textbf{EKF Nav.}
\\
\midrule

Kassas et al. (2024)~\cite{kassas2024adastra}
& \checkmark & -- & -- & -- & -- & \checkmark \\

Shahcheraghi et al. (2025)~\cite{shahcheraghi2025acq}
& \checkmark & -- & -- & -- & -- & \checkmark \\

Neinavaie et al. (2022)~\cite{neinavaie2022differential}
& \checkmark & -- & -- & -- & -- & \checkmark \\

Abdallah and Kassas (2021)~\cite{abdallah2021multipath}
& \checkmark & \checkmark & -- & -- & \checkmark & -- \\

Chang et al. (2024)~\cite{chang2024joint}
& -- & \checkmark & \checkmark & -- & \checkmark & -- \\

Li et al. (2024)~\cite{sblnlos2024}
& \checkmark & -- & -- & -- & -- & -- \\

Shan et al. (2025)~\cite{shan2025offgrid}
& -- & -- & \checkmark & -- & \checkmark  & -- \\

He et al. (2026)~\cite{he2026offgrid}
& \checkmark & -- & \checkmark & -- & -- & -- \\

\midrule
\textbf{Proposed framework}
& \checkmark & \checkmark & \checkmark & \checkmark & \checkmark & \checkmark \\
\bottomrule
\end{tabular}%
}
\end{table*} 
The simultaneous tracking and navigation (STAN) framework jointly estimates receiver and LEO satellite states using an EKF to account for ephemeris and clock uncertainties, and has been experimentally validated with multiple LEO constellations in GNSS-denied environments~\cite{kassas2024adastra}. Differential LEO-SoOP techniques use a known-position reference receiver to provide measurements from the same LEO satellites observed by a rover, with analytical studies showing improved positioning accuracy through differential Doppler processing~\cite{neinavaie2022differential}. Differential STAN (DSTAN) has also been proposed to further improve positioning performance~\cite{kassas2023lookstars}.

These studies have made significant progress in addressing the first three challenges; however, urban environments with multipath and non-line-of-sight (NLOS), as illustrated in Fig.~\ref{fig:enter-label}, remain challenging for LEO-SoOP navigation.
 Recent studies have incorporated LOS and NLOS into wideband LEO signal models~\cite{ma2025leo}, while the larger Doppler separation between direct and reflected components at higher carrier frequencies may offer opportunities for multipath mitigation~\cite{zhang2025multipath}. Nevertheless, Doppler separation does not eliminate multipath, and closely spaced LOS and NLOS may remain unresolved when their delay separation approaches or falls below the effective delay resolution, making reliable delay--Doppler estimation difficult~\cite{zhang2025multipath,abdallah2021multipath}. This challenge is further compounded by non-ideal noise, including non-Gaussian, impulsive, and temporally correlated disturbances. 
 
 Conventional GLRT-based acquisition using delay--Doppler cross-ambiguity function (CAF) faces two key limitations: its performance can degrade under mismatched noise statistics~\cite{kay1998detection,sengupta1990glrt}, and overlapping LOS and multipath correlation responses can shift the detected peak, producing biased delay--Doppler estimates~\cite{abdallah2021multipath}. Although robust GLRT formulations can mitigate non-Gaussian and correlated disturbances~\cite{sengupta1990glrt,daskovic2020robust}, they do not resolve the ambiguity between closely spaced multipath components. Consequently, delay--Doppler biases may persist even at high SNR, motivating high-resolution methods capable of separating LOS and multipath under realistic urban noise conditions.

Sparse recovery and compressed sensing (CS) techniques have been widely investigated for sparse multipath estimation by exploiting the limited number of propagation paths. Representative approaches include greedy methods such as orthogonal matching pursuit (OMP)~\cite{tropp2007omp}, $\ell_1$-norm minimization~\cite{chen1998atomic}, and gridless methods such as atomic norm minimization~\cite{chandrasekaran2012atomic}. Although these methods can achieve effective sparse recovery, their performance may be affected by dictionary mismatch, closely spaced components, and model or regularization choices. sparse Bayesian learning (SBL) provides an alternative probabilistic framework based on automatic relevance determination (ARD)~\cite{tipping2001,wipf2004}, enabling adaptive identification and suppression of insignificant components while retaining dominant propagation paths. SBL has been applied to GNSS multipath mitigation through joint spatial, polarization, and temporal estimation~\cite{chang2024joint} and joint angle--delay estimation~\cite{chang2021joint}, as well as to NLOS channel estimation in LEO satellite communications~\cite{sblnlos2024}.

Conventional SBL, however, represents delay-Doppler on a fixed discrete grid, and off-grid parameters can therefore cause estimation errors. Off-grid SBL addresses this limitation by refining continuous-valued parameters beyond the predefined grid~\cite{wipf2010offgrid,shan2025offgrid}. Recent studies have investigated two-dimensional off-grid SBL for delay--Doppler estimation in LEO satellite--ground links~\cite{he2026offgrid}, Doppler estimation in OTFS systems~\cite{shan2025offgrid}, AFDM-ISAC target sensing~\cite{liu2025afdm}, and radar imaging and localization~\cite{pote2023offgrid}. These studies demonstrate the potential of SBL-based methods for high-resolution sparse parameter estimation. However, as summarized in Table~\ref{tab:literature_comparison}, most existing studies treat sparse channel estimation and off-grid parameter recovery separately and do not jointly address closely spaced LOS and NLOS components and non-ideal noise in LEO-SoOP navigation.

\begin{figure}
    \centering\includegraphics[width=1.02\linewidth]{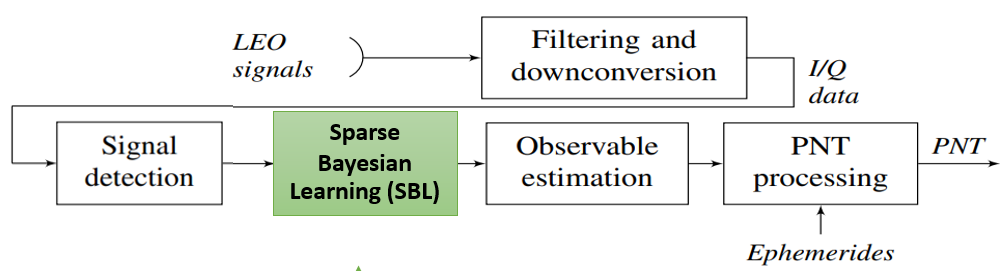}
    \caption{ The proposed LEO-SoOP receiver architecture.}
    \label{fig:enter-label1}
\end{figure}

To the best of our knowledge, no existing study has jointly addressed LOS and NLOS separation in the delay--Doppler domain, robust measurement refinement under non-Gaussian and temporally correlated noise, and reliability-aware integration into LEO-SoOP navigation. To address this gap, this paper proposes a hybrid robust GLRT and off-grid SBL framework. As illustrated in Fig.~\ref{fig:enter-label1}, the framework integrates (i) GLRT-based signal acquisition, (ii) off-grid SBL refinement for multipath separation and continuous delay--Doppler estimation, and (iii) SBL-aided robust navigation.

The main contributions are summarized as follows:

\begin{itemize}
\item We investigate the impact of NLOS, non-Gaussian, and temporally correlated disturbances on GLRT-based acquisition in urban LEO-SoOP, highlighting its limitations under realistic conditions.

\item We develop a GLRT-assisted off-grid SBL framework for joint LOS and NLOS separation and accurate delay--Doppler estimation, bridging GLRT acquisition with continuous off-grid refinement.

\item We introduce a hierarchical Bayesian model with ARD-based sparsity priors and a Student's-$t$ noise to identify relevant propagation paths and remain robust to heavy-tailed disturbances and outliers.

\item We integrate SBL-derived reliability information into the EKF through adaptive covariance weighting, allowing the filter to downweight unreliable or multipath contaminated measurements and improve navigation robustness.

\item Simulations under non-ideal noise and severe urban multipath show that the proposed framework outperforms GLRT in delay--Doppler accuracy, LOS and NLOS separation, and navigation performance.

\end{itemize}

The rest of this paper is organized as follows. Section~\ref{sec:signal_channel_model}
presents the signal model. Section~\ref{sec:glrt_analysis}
analyzes GLRT bias under urban multipath. Section~\ref{sec:framework}
describes the proposed off-grid SBL framework. Section~\ref{sec:crlb}
presents the Cram\'er--Rao bound analysis. Section~\ref{sec:nav}
details the SBL-aided robust EKF navigation framework. Section~\ref{sec:results}
presents the simulation results. Section~\ref{sec:conclusion} concludes.

\section{Signal and Channel Model}
\label{sec:signal_channel_model}

The received baseband signal is modeled as
\begin{equation}
r[n] = \sum_{i=1}^{K} \sum_{p=1}^{P_i} a_{i,p} \, s_i(nT_s - \tau_{i,p}) \, e^{j2\pi f_{D_{i,p}} T_s n} + w[n],
\label{eq:received_signal}
\end{equation}
where $r[n]$ is the received complex baseband sample at discrete time index $n$, $K$ is the number of visible LEO satellites, and $P_i$ is the number of propagation paths associated with the $i$-th satellite. The parameter $a_{i,p} \in \mathbb{C}$ is the complex gain of the $p$th propagation path, $s_i$ represents samples of the transmitted signal from the $i$th LEO satellite, $\tau_{i,p}$ is the propagation delay, and $f_{D_{i,p}} = \nu_{i,p}$ is the corresponding Doppler shift. Furthermore, $T_s$ is the sampling interval, and $w[n]$ represents additive receiver noise. Here, the noise for the $i$th satellite is modeled as an AR(1) process driven by
Student's-$t$ innovations, capturing both temporal correlation and
non-Gaussian behavior in the urban environment~\cite{christmas2011robust}:
\begin{equation}
w_i[n]
=
\rho w_i[n-1]
+
\sqrt{1-\rho^2}\,\varepsilon_i[n],
\label{eq:ar1_noise}
\end{equation}
where $\varepsilon_i[n]
\sim
t_\nu(0,\sigma_\varepsilon^2),
\label{eq:student_t_innovation}$
with $\rho$ denoting the AR(1) correlation coefficient and $\nu$
denoting the degrees of freedom of the Student's $t$ distribution.

Under a sufficiently short coherent processing interval (CPI), and after Doppler-rate compensation, the propagation delay--Doppler shift are assumed approximately constant over the CPI~\cite{shahcheraghi2025acq}. The resulting $N$ sample observation vector for the $i$th satellite is 
\begin{equation}
\mathbf{y}_i = \sum_{p=1}^{P_i} a_{i,p} \, \mathbf{d}_i(\tau_{i,p}, \nu_{i,p}) + \mathbf{w}_i,
\label{eq:cpi_model}
\end{equation}
where $\mathbf{w}_i \in \mathbb{C}^N$ is the additive noise vector, and $\mathbf{d}_i(\tau_{i,p}, \nu_{i,p})$ is the delay--Doppler atom, with $n$th element
\begin{equation}
\left[ \mathbf{d}_i(\tau, \nu) \right]_n = \hat{s}_i \bigl( (nT_s - \tau) \bmod \hat{T}_{0,i} \bigr) \, e^{j2\pi \nu n T_s}.
\label{eq:atom}
\end{equation}
Here, $\hat{s}_i(\cdot)$ denotes the estimate of the transmitted signal, and $\hat{T}_{0,i}$ is the estimated repetition period of the transmitted signal by the $i$th LEO satellite. The continuous delay--Doppler domain is discretized into a predefined grid to obtain a sparse multipath representation.\begin{equation}
\mathcal{G}_i = \left\{ (\tau_m, \nu_\ell) : m = 1, \dots, M_\tau, \; \ell = 1, \dots, M_\nu \right\},
\label{eq:grid}
\end{equation}
where $\mathcal{G}_i$ denotes the delay--Doppler search grid for the $i$th satellite, with $M_\tau$ and $M_\nu$ delay and Doppler points, respectively. Each grid point is indexed by $q=1,\dots,M$, where $M=M_\tau M_\nu$. The delay--Doppler atom is 
\begin{equation}
\boldsymbol{\phi}_{i,q} = \mathbf{d}_i(\tau_m, \nu_\ell).
\label{eq:atom_grid}
\end{equation}
Since the number of physical propagation paths is much smaller than the number of candidate delay-Doppler grid points, $P_i \ll M$, only a small number of dictionary atoms are active. Therefore, the CPI observation can be represented by the sparse linear model
\begin{equation}
\label{eq:sparse_model}
  \mathbf{y}_i \approx \boldsymbol\Phi_i\mathbf{x}_i
  + \mathbf{w}_i,\quad
  \mathbf{x}_i \in \mathbb{C}^M,\;
  \|\mathbf{x}_i\|_0 = P_i,
\end{equation}
 where $\mathbf{x}_i$ is a sparse vector whose nonzero entries represent the active paths and their delay-Doppler grid points and $\boldsymbol{\Phi}_i$ is delay--Doppler dictionary.

\section{GLRT Bias Analysis in urban environments}
\label{sec:glrt_analysis}

\subsection{The conventional GLRT Estimator}
The conventional acquisition of LEO SoOP signals is commonly performed using
a GLRT-based on the delay--Doppler CAF~\cite{shahcheraghi2025acq}. The delay--Doppler estimate is 
\begin{equation}
(\hat{\tau}_{\mathrm{GLRT}},\hat{\nu}_{\mathrm{GLRT}})
=
\arg\max_{\tau,\nu}
\frac{
\left|
\mathbf{d}^{H}(\tau,\nu)\mathbf{y}
\right|^2
}{
\left\|\mathbf{d}(\tau,\nu)\right\|_2^2
},
\label{eq:glrt_estimator_norm}
\end{equation}

A dominant LOS component and additive white Gaussian noise (AWGN) allow
the conventional GLRT to provide efficient delay--Doppler estimates with
low computational complexity~\cite{kay1998detection}. However, in dense
urban environments, reflected NLOS components and non-ideal noise can
violate these assumptions and degrade estimation performance.

\subsection{ GLRT performance in urban environments}

In the presence of multipath \eqref{eq:received_signal}, the CAF-based GLRT still searches for a single dominant delay--Doppler
peak. However, the correlation responses of
the LOS and NLOS components may overlap, causing the detected peak to
deviate from the true LOS delay--Doppler~\cite{abdallah2021multipath}. The resulting shift in the detected peak
displacement can be characterized by the following power-weighted
approximation of the GLRT delay estimate:
\begin{equation}
\hat{\tau}_{\mathrm{GLRT}}
\approx
\tau_0
+
\frac{
\sum_{p=1}^{P}|a_p|^2\Delta\tau_p
}{
|a_0|^2+\sum_{p=1}^{P}|a_p|^2
}.
\label{eq:glrt_delay_bias}
\end{equation}
where $a_0$ and $\tau_0$ denote the complex amplitude and delay of the LOS, respectively. The parameter $P$
denotes the number of NLOS components, while $a_p$ and $\tau_p$ represent the complex amplitude and delay of
the $p$th NLOS component, respectively.  $\Delta\tau_p=\tau_p-\tau_0$ is the excess delay of the
$p$th NLOS component. The derivation is provided in Appendix~\ref{app:bias_proof}.This bias arises from the mismatch between the single-path model assumed by the conventional GLRT and the actual multipath channel, rather than from random thermal noise. Therefore, increasing the SNR or coherent integration time can reduce the random estimation error but cannot eliminate the multipath-induced bias. At sufficiently high SNR, this bias can become the dominant source of delay-estimation error. 

Urban LEO-SoOP signals are affected by
temporally correlated and non-Gaussian noise, which can further degrade
delay--Doppler estimation. GLRT-based detection has been studied under
such non-ideal noise conditions~\cite{sengupta1990glrt,daskovic2020robust}. The Robust GLRT whitens the received signal to handle temporally correlated disturbances and reduces the effect of large-amplitude samples caused by non-Gaussian disturbances.

\section{SBL-Based Multipath Mitigation Framework}
\label{sec:framework}
The GLRT-based acquisition provides a computationally efficient initial estimate but can degrade in urban LEO-SoOP environments due to its single-path assumption and sensitivity to non-ideal noise statistics~\cite{kay1998detection}. An off-grid SBL framework is therefore developed for high-resolution estimation and multipath separation by representing the received signal as a sparse combination of delay--Doppler components~\cite{tipping2001,wipf2004,wipf2010offgrid}. This enables the significant propagation components to be inferred directly from the received signal.

The proposed framework consists of three stages. First, a compact delay--Doppler dictionary is constructed using TLE-based orbital information and GLRT-assisted acquisition to restrict the search region and reduce computational cost. Second, an off-grid sparse representation accounts for continuous delay--Doppler locations and grid mismatch. Finally, an expectation--maximization (EM) SBL procedure identifies the active propagation components and continuously refines their delay--Doppler parameters.

\subsection{TLE and GLRT-Aided Off-Grid Dictionary Design}
\label{subsubsec:tle_glrt_dict}
The first step constructs a compact delay--Doppler dictionary focused on the most plausible propagation region, reducing the computational cost of sparse recovery. TLE-based orbital information provides initial delay--Doppler estimates and defines an uncertainty region accounting for orbital prediction, receiver-state, and propagation-model errors. The GLRT-based acquisition then estimates the dominant component. The dictionary construction procedure is summarized in Algorithm~\ref{alg:tle_glrt_offgrid}. The TLE-based predicted delay and Doppler for satellite $i$ at time step $k$ are obtained as
\begin{subequations}
\begin{align}
\hat{\tau}_i^{\mathrm{TLE}}(k)
&=
\frac{
\left\| \hat{\mathbf{r}}_{r}^{\mathrm{pred}}(k)
-
\hat{\mathbf{r}}_{\mathrm{leo},i}^{\mathrm{TLE}}(k)
\right\|_2
}{c},
\\[2mm]
\hat{\nu}_i^{\mathrm{TLE}}(k)
&=
-\frac{f_{c_i}}{c}
\,
\frac{
\left[
\hat{\dot{\mathbf{r}}}_{r}^{\mathrm{pred}}(k)
-
\hat{\dot{\mathbf{r}}}_{\mathrm{leo},i}^{\mathrm{TLE}}(k)
\right]^T
\left[
\hat{\mathbf{r}}_{r}^{\mathrm{pred}}(k)
-
\hat{\mathbf{r}}_{\mathrm{leo},i}^{\mathrm{TLE}}(k)
\right]
}{
\left\|
\hat{\mathbf{r}}_{r}^{\mathrm{pred}}(k)
-
\hat{\mathbf{r}}_{\mathrm{leo},i}^{\mathrm{TLE}}(k)
\right\|_2
}.
\end{align}
\label{eq:tle_pred}
\end{subequations}
where $f_{c_i}$ is carrier frequency and $c$ is speed of light.
A coarse receiver state estimate with initial position and velocity errors is used to construct the first TLE-informed delay--Doppler search region. After initialization, the predicted receiver position $\hat{\mathbf r}_{r}^{\mathrm{pred}}(k)$ and velocity $\hat{\dot{\mathbf r}}_{r}^{\mathrm{pred}}(k)$ are obtained by propagating the previous navigation state to the current epoch. Satellite position $\mathbf r_{\mathrm{leo},i}^{\mathrm{TLE}}(k)$ and velocity $\dot{\mathbf r}_{\mathrm{leo},i}^{\mathrm{TLE}}(k)$ are obtained from TLE-based SGP4 propagation~\cite{shahcheraghi2025acq,khairallah2024ephemeris}.

The
GLRT is then applied to obtain the coarse
delay--Doppler estimate
$(\hat{\tau}_i^{\mathrm{GLRT}},\hat{\nu}_i^{\mathrm{GLRT}})$, as defined
in \eqref{eq:glrt_estimator_norm}. Based on these estimates, the fine delay--Doppler search region is
defined around the GLRT result and constrained by TLE-informed
uncertainty region:
\begin{equation}
{\scriptstyle
\mathcal{G}_i^{\mathrm{fine}}
=
\left\{
(\tau_g,\nu_g):
\begin{array}{l}
\tau_g \in
\left[
\hat{\tau}_i^{\mathrm{GLRT}}\pm\Delta\tau_{\mathrm{fine}}
\right]
\cap
\left[
\hat{\tau}_i^{\mathrm{TLE}}\pm\Delta\tau_{\max}
\right],
\\[1mm]
\nu_g \in
\left[
\hat{\nu}_i^{\mathrm{GLRT}}\pm\Delta\nu_{\mathrm{fine}}
\right]
\cap
\left[
\hat{\nu}_i^{\mathrm{TLE}}\pm\Delta\nu_{\max}
\right]
\end{array}
\right\}}
\label{eq:fine_grid}
\end{equation}
where $\Delta\tau_{\max}$ and $\Delta\nu_{\max}$ define the uncertainty
bounds associated with the TLE-based delay--Doppler predictions,
respectively, accounting for TLE and SGP4 propagation errors and
uncertainty in the predicted receiver state. The parameters
$\Delta\tau_{\mathrm{fine}}$ and $\Delta\nu_{\mathrm{fine}}$ define the
local refinement half-widths around the GLRT estimate. 

Finally, the sparse dictionary $\boldsymbol{\Phi}_i$ is constructed by
evaluating the delay--Doppler atoms over the fine search region
$\mathcal{G}_i^{\mathrm{fine}}$:
\begin{equation}
\boldsymbol{\Phi}_i
=
\left[
\mathbf{d}_i(\tau_1,\nu_1),
\mathbf{d}_i(\tau_2,\nu_2),
\dots,
\mathbf{d}_i(\tau_{G_i^{\text{fine}}}, \nu_{G_i^{\text{fine}}})
\right].
\label{eq:fine_dictionary}
\end{equation}

The dictionary contains $G_i^{\mathrm{fine}}=\left|\mathcal{G}_i^{\mathrm{fine}}\right|$ delay--Doppler atoms for satellite $i$. Although it is restricted to a localized region around the coarse acquisition estimate, it is constructed on a discrete delay--Doppler grid, whereas the actual delay--Doppler of each propagation component are continuous-valued. Thus, a propagation path may lie between adjacent grid points, and representing it by the nearest grid atom introduces grid mismatch and estimation errors. A finer grid can reduce this error but increases the dictionary size and computational complexity. The proposed method allows the actual delay and Doppler to lie between
the predefined grid points~\cite{shan2025offgrid,liu2025afdm}, given by
\begin{equation}
\tau_{i,p} = \tau_{g,j_p} + \delta\tau_{i,j_p},
\qquad
\nu_{i,p} = \nu_{g,j_p} + \delta\nu_{i,j_p},
\label{eq:offgrid_parameters}
\end{equation}
where $\tau_{g,j_p}$ and $\nu_{g,j_p}$ denote the delay and Doppler
coordinates of the $j_p$th active grid atom, respectively, while
$\delta\tau_{i,j_p}$ and $\delta\nu_{i,j_p}$ denote the
continuous offsets from this grid point. The off-grid offsets are constrained
to remain within their corresponding grid cells, i.e.,
$|\delta\tau_{i,j_p}|\leq\Delta\tau_g/2$ and
$|\delta\nu_{i,j_p}|\leq\Delta\nu_g/2$. These offsets are incorporated
into the dictionary using a first-order Taylor expansion around each
grid point, yielding
\begin{equation}
\boldsymbol{\Phi}_{\mathrm{eff},i}
=
\boldsymbol{\Phi}_i
+
\boldsymbol{\Psi}^{\tau}_i
\operatorname{diag}(\boldsymbol{\delta\tau}_i)
+
\boldsymbol{\Psi}^{\nu}_i
\operatorname{diag}(\boldsymbol{\delta\nu}_i),
\label{eq:eff_dictionary}
\end{equation}
where $\boldsymbol{\Psi}^{\tau}_i$ and $\boldsymbol{\Psi}^{\nu}_i$ contain the derivatives of the dictionary atoms with respect to delay and Doppler, respectively. The Doppler and delay derivatives for $j$th atom are
\begin{equation}
\begin{aligned}
\left[\boldsymbol{\Psi}^{\nu}_{i,j}\right]_n
&=
j2\pi nT_s \left[\boldsymbol{\Phi}_{i,j}\right]_n,\\
\left[\boldsymbol{\Psi}^{\tau}_{i,j}\right]_n
&=
e^{j2\pi\nu_{g,j}nT_s}
\left.
\frac{\partial
\hat{s}_i\!\left(
(nT_s-\tau)\bmod \hat{T}_{0,i}
\right)}
{\partial \tau}
\right|_{\tau=\tau_{g,j}} .
\end{aligned}
\label{eq:derivatives}
\end{equation}

The proposed approach uses whitening to account for the temporal correlation in the noise. Accordingly, the whitened dictionary and observation model are written as
\begin{equation}
\boldsymbol{\Phi}_{w,\mathrm{eff},i}
=
\mathbf W_i \, \boldsymbol{\Phi}_{\mathrm{eff},i}, \qquad
\mathbf{y}_{w,i}
\approx
\boldsymbol{\Phi}_{w,\mathrm{eff},i} \, \mathbf{x}_i
+
\mathbf{w}_{w,i},
\end{equation}
where $\mathbf W_i$ denotes the whitening matrix for the $i$th signal. The AR(1) correlation is removed, leaving temporally independent Student's-$t$ innovations. The SBL model is also made robust to impulsive and non-Gaussian disturbances
by modeling the measurement noise as a Gaussian scale mixture:
\begin{equation}
\mathbf{w}_{w,i}\mid \boldsymbol{\lambda}_i
\sim
\mathcal{CN}\bigl(\mathbf{0},\, \boldsymbol{\Lambda}_i^{-1}\bigr),
\label{eq:gsm_noise}
\end{equation}
where $\boldsymbol{\Lambda}_i =
\operatorname{diag}(\lambda_{i,1},\dots,\lambda_{i,N})$ is the noise
precision matrix, whose precision variables are estimated from the data
rather than marginalized as additional latent variables, and follow a
Gamma distribution.

\begin{algorithm}[t]
\caption{ Off-Grid Dictionary Design}
\label{alg:tle_glrt_offgrid}
\begin{algorithmic}[1]
\REQUIRE $\mathbf{y}_i$, TLE data,  predicted receiver state,
$\Delta\tau_{\max}$, $\Delta\nu_{\max}$,
$\Delta\tau_{\mathrm{fine}}$, $\Delta\nu_{\mathrm{fine}}$

\ENSURE Effective dictionary $\boldsymbol{\Phi}_{\mathrm{eff},i}$

\STATE Compute TLE-based delay--Doppler estimates \eqref{eq:tle_pred}
 
\STATE Compute GLRT-based delay--Doppler estimates \eqref{eq:glrt_estimator_norm}

\STATE Construct the fine delay--Doppler grid
$\mathcal{G}_i^{\mathrm{fine}}$ \eqref{eq:fine_grid}.

\STATE Construct the discrete dictionary
$\boldsymbol{\Phi}_i$ within $\mathcal{G}_i^{\mathrm{fine}}$ \eqref{eq:fine_dictionary}

\STATE Initialize the off-grid offsets
$\boldsymbol{\delta\tau}_i$ and $\boldsymbol{\delta\nu}_i$ to zero

\STATE Construct the effective dictionary
$\boldsymbol{\Phi}_{\mathrm{eff},i}$ using
\eqref{eq:eff_dictionary}

\RETURN $\boldsymbol{\Phi}_{\mathrm{eff},i}$
\end{algorithmic}
\end{algorithm}

\subsection{Hierarchical SBL Formulation}
\label{subsubsec:bayesian_model}

After constructing the delay--Doppler dictionary and effective model, the next step is to identify the dictionary atoms associated with propagation paths present in the received signal. To this end, each dictionary atom is assigned a complex coefficient $x_{i,j}$ in SBL formulation, representing the contribution of a delay--Doppler component to received signal. An independent Bayesian prior is assigned to each coefficient to promote sparsity, retaining significant propagation components while suppressing coefficients associated with irrelevant atoms. The prior distribution is 
\begin{equation}
p(\mathbf{x}_i \mid \boldsymbol{\gamma}_i)
=
\prod_{j=1}^{G_i^{\mathrm{fine}}}
\mathcal{CN}\bigl(x_{i,j}\mid 0,\gamma_{i,j}\bigr).
\label{eq:sparse_prior}
\end{equation}
where $\boldsymbol{\gamma}_i =
[\gamma_{i,1},\ldots,\gamma_{i,G_i^{\mathrm{fine}}}]^T$ denotes the vector of variance hyperparameters, with $\gamma_{i,j}$ controlling the variance of $x_{i,j}$.
The ARD mechanism learns these hyperparameters from
the received signal, driving those associated with irrelevant components
toward zero while retaining non-negligible values for relevant
propagation components. Consequently, the active propagation paths are identified directly from
data without specifying their number in advance~\cite{tipping2001,wipf2004}.

The hyperparameters, including the SBL variance hyperparameters,
noise precision, and off-grid delay and Doppler corrections, are estimated by maximizing the marginal likelihood (Type-II maximum
likelihood (ML)):
\begin{equation}
\max_{\boldsymbol{\gamma}_i, \boldsymbol{\lambda}_i, \boldsymbol{\delta\tau}_i, \boldsymbol{\delta\nu}_i}
\log
\int
p(\mathbf{y}_{w,i} \mid \mathbf{x}_i, \boldsymbol{\lambda}_i, \boldsymbol{\delta\tau}_i, \boldsymbol{\delta\nu}_i)
\,
p(\mathbf{x}_i \mid \boldsymbol{\gamma}_i)
\,
d\mathbf{x}_i .
\label{eq:typeII_ml}
\end{equation}

The Type-II ML problem is solved using the EM algorithm~\cite{wipf2004},
where the sparse coefficients $\mathbf{x}_i$ are treated as latent
variables and the SBL hyperparameters $\boldsymbol{\gamma}_i$,
noise precision parameters $\boldsymbol{\lambda}_i$, and off-grid
corrections $\boldsymbol{\delta\tau}_i$ and $\boldsymbol{\delta\nu}_i$
are iteratively updated to identify active propagation components,
model non-Gaussian and temporally correlated noise, and refine their
delay--Doppler locations. The off-grid EM-SBL refinement procedure is
summarized in Algorithm~\ref{alg:sbl_refinement}.

The EM procedure starts from a sparse representation obtained from the reduced delay--Doppler dictionary. OMP provides the initialization by greedily selecting atoms that best match the residual~\cite{tropp2007omp}. FMLM~\cite{tipping2003fast} then sequentially adds, removes, or re-estimates atoms based on their marginal-likelihood contribution, retaining relevant active atoms for subsequent EM-SBL inference.

\subsubsection{Posterior Estimation (E-step)}
\label{subsubsec:e_step}

  Hyperparameters and effective dictionary define the posterior
distribution of the sparse coefficient vector \begin{equation}
p(\mathbf{x}_i \mid \mathbf{y}_{w,i})
=
\mathcal{CN}\bigl(\boldsymbol{\mu}_i^{(t)},\,\boldsymbol{\Sigma}_i^{(t)}\bigr),
\label{eq:posterior}
\end{equation}
where the posterior covariance and mean are given by
\begin{equation}
\boldsymbol{\Sigma}_i^{(t)}
=
\left(
\boldsymbol{\Phi}_{\mathrm{eff},{w,i}}^{(t)H}
\boldsymbol{\Lambda}_i^{(t)}
\boldsymbol{\Phi}_{\mathrm{eff},{w,i}}^{(t)}
+
\operatorname{diag}(\boldsymbol{\gamma}_i^{(t)})^{-1}
\right)^{-1},
\label{eq:e_step_cov}
\end{equation}
\begin{equation}
\boldsymbol{\mu}_i^{(t)}
=
\boldsymbol{\Sigma}_i^{(t)}
\boldsymbol{\Phi}_{\mathrm{eff},{w,i}}^{(t)H}
\boldsymbol{\Lambda}_i^{(t)}
\mathbf{y}_{w,i}.
\label{eq:e_step_mean}
\end{equation}
The posterior mean estimates contribution of each dictionary atom,
while the posterior covariance captures the uncertainty.

Finally, delay-Doppler corrections are refined using 
posterior estimates and current noise precision weights:
\begin{equation}
\delta\tau_{i,j}^{(t+1)}
=
\frac{
\operatorname{Re}
\left[
\left( \dot{\boldsymbol{\Phi}}_{i,j}^{\tau} \right)^H
\boldsymbol{\Lambda}_i^{(t+1)}
\mathbf{r}_i^{(t)}
\mu_{i,j}^{(t)*}
\right]
}{
\left( \dot{\boldsymbol{\Phi}}_{i,j}^{\tau} \right)^H
\boldsymbol{\Lambda}_i^{(t+1)}
\dot{\boldsymbol{\Phi}}_{i,j}^{\tau}
\,
\left| \mu_{i,j}^{(t)} \right|^2
+ \varepsilon
}.
\label{eq:m_step_deltatau}
\end{equation}

\begin{equation}
\delta\nu_{i,j}^{(t+1)}
=
\frac{
\operatorname{Re}
\left[
\left( \dot{\boldsymbol{\Phi}}_{i,j}^{\nu} \right)^H
\boldsymbol{\Lambda}_i^{(t+1)}
\mathbf{r}_i^{(t)}
\mu_{i,j}^{(t)*}
\right]
}{
\left( \dot{\boldsymbol{\Phi}}_{i,j}^{\nu} \right)^H
\boldsymbol{\Lambda}_i^{(t+1)}
\dot{\boldsymbol{\Phi}}_{i,j}^{\nu}
\,
\left| \mu_{i,j}^{(t)} \right|^2
+ \varepsilon
}.
\label{eq:m_step_deltanu}
\end{equation}
where $\varepsilon>0$ is a small regularization constant used to ensure numerical stability. The EM algorithm iteratively updates the hyperparameters until convergence. Convergence is declared when the relative change in the sparsity-controlling hyperparameters $\gamma_i$ falls below $10^{-5}$ after at least 10 iterations. Once convergence is reached, the final
hyperparameters are used to identify the dictionary atoms associated
with significant paths.  Each identified path is
characterized by its estimated delay--Doppler pair. LOS is selected from these significant paths by
comparing their estimated delay--Doppler with TLE
predictions. 
Among the candidate paths that satisfy both delay--Doppler
constraints, the path with smallest estimated delay is selected as LOS.

\begin{algorithm}[t]
\caption{ Off-Grid EM-SBL Multipath Refinement}
\label{alg:sbl_refinement}
\begin{algorithmic}[1]

\REQUIRE $\mathbf{y}_i$, TLE-based delay--Doppler prediction, reduced dictionary $\boldsymbol{\Phi}_i$

\ENSURE LOS delay--Doppler estimate and
$\bar{\lambda}_i^{\mathrm{SBL}}$

\STATE Initial sparse coefficient vector
$\hat{\mathbf{x}}_i$ using OMP or FMLM.

\STATE Initialize the posterior mean as
$\boldsymbol{\mu}_i^{(0)}=\hat{\mathbf{x}}_i$ and the remaining hyperparameters are initialized as
\\$\gamma_{i,j}^{(0)} = 1,
\lambda_{i,n}^{(0)} = 1,
\delta\tau_{i,j}^{(0)} = 0,
\delta\nu_{i,j}^{(0)} = 0.$

\WHILE{the convergence criterion is not satisfied}

    \STATE Update the effective dictionary.

    \STATE \textbf{E-step:} 
    Compute
$\boldsymbol{\Sigma}_i^{(t)}$ and
    $\boldsymbol{\mu}_i^{(t)}$ 
    \eqref{eq:e_step_cov}/\eqref{eq:e_step_mean}.

    \STATE \textbf{M-step:} 
    Update
$\boldsymbol{\gamma}_i^{(t+1)}$  \eqref{eq:m_step_gamma}.

   Update the noise precision
    $\boldsymbol{\lambda}_i^{(t+1)}$ \eqref{eq:m_step_lambda}.

     Update the off-grid delay--Doppler corrections
    \eqref{eq:m_step_deltatau}/\eqref{eq:m_step_deltanu}.

\ENDWHILE

\STATE Identify the significant propagation paths using the final
ARD hyperparameters.

\STATE Form TLE-consistent candidate set, select the LOS.

\STATE Compute SBL reliability indicator
$\bar{\lambda}_i^{\mathrm{SBL}}$ \eqref{eq:lambda_sbl} and return to navigation layer.

\end{algorithmic}
\end{algorithm}
\subsubsection{Hyperparameter Updates (M-Step)}
\label{subsubsec:m_step}

The E-step is used to update the hyperparameters:
\begin{equation}
\gamma_{i,j}^{(t+1)}
=
\left|\mu_{i,j}^{(t)}\right|^2
+
\Sigma_{i,jj}^{(t)}.
\label{eq:m_step_gamma}
\end{equation}

The noise precision variables are then updated as

\resizebox{0.95\columnwidth}{!}{%
\begin{minipage}{\columnwidth}
\begin{equation}
\lambda_{i,n}^{(t+1)}
=
\frac{
a+1
}{
b+
\left|
y_{i,n}-
\left(
\boldsymbol{\Phi}_{\mathrm{eff},{w,i}}^{(t)}
\boldsymbol{\mu}_i^{(t)}
\right)_n
\right|^2
+
\left[
\boldsymbol{\Phi}_{\mathrm{eff},{w,i}}^{(t)}
\boldsymbol{\Sigma}_i^{(t)}
\boldsymbol{\Phi}_{\mathrm{eff},{w,i}}^{(t)H}
\right]_{n,n}
}.
\label{eq:m_step_lambda}
\end{equation}
\end{minipage}%
}
\section{Cram\'er--Rao Bound Analysis}
\label{sec:crlb}

In this section, the Cram\'er--Rao lower bound (CRLB) is evaluated for the proposed framework.

\subsubsection{Single-Path CRLB}

The CRLBs for the single-path LOS delay and Doppler under AWGN are
given by~\cite{kay1993}.
\begin{equation}
\CRLB_{\tau_0}^{\mathrm{Single-Path}}
=
\frac{1}
{8\pi^2\beta^2\,\mathrm{SNR}\,N},
\label{eq:crlb_sp}
\end{equation}
\begin{equation}
\CRLB_{\nu_0}^{\mathrm{Single-Path}}
=
\frac{6}
{(2\pi)^2T_s^2\,
\mathrm{SNR}\,
N(N^2-1)},
\label{eq:crlb_nu_sp}
\end{equation}
where
$\mathrm{SNR}=|a_0|^2/\sigma^2$,
$\beta$ is RMS bandwidth, and $T_s$ is
sampling interval.

\subsubsection{Multipath CRLB}
In an urban environment, multipath introduces additional delayed and Doppler-shifted NLOS components that act as nuisance parameters in LOS estimation. Let $\boldsymbol{\eta}$ collect these nuisance parameters, while the path amplitudes and Doppler shifts associated with the unknown multipath components are assumed to be fixed. The Fisher Information Matrix (FIM) is
\begin{equation}
\mathbf J=
\begin{bmatrix}
J_{\tau_0\tau_0} &
\mathbf J_{\tau_0\eta}\\
\mathbf J_{\eta\tau_0} &
\mathbf J_{\eta\eta}
\end{bmatrix}.
\label{eq:fim_partition}
\end{equation}
where $J_{\tau_0\tau_0}$ is the LOS-delay information,
$\mathbf{J}_{\tau_0\eta}$ and $\mathbf{J}_{\eta\tau_0}$ describe the
LOS--multipath coupling, and $\mathbf{J}_{\eta\eta}$ contains the
information associated with the multipath parameters.

The Schur complement accounts for  information loss~\cite{kay1993}, leading to the following
CRLB for LOS delay:
\begin{equation}
\CRLB_{\tau_0}^{\mathrm{Multipath}}
=
\left[
J_{\tau_0\tau_0}
-
\mathbf J_{\tau_0\eta}
\mathbf J_{\eta\eta}^{-1}
\mathbf J_{\eta\tau_0}
\right]^{-1}.
\label{eq:crlb_mp}
\end{equation}

The second term represents the information loss caused by the coupling
between the LOS delay and the unknown multipath parameters. This coupling
becomes stronger when the paths are closely spaced in delay relative to
the inverse signal bandwidth, making the multipath components harder to
distinguish and potentially degrading LOS-delay estimation. Thus, the
impact of multipath depends on the path characteristics and waveform and
cannot be captured by a single multiplicative factor. Meanwhile, this Formulation provides a theoretical benchmark for the
proposed SBL estimator. By exploiting the sparse structure of the
delay--Doppler domain and refining the path parameters off-grid, SBL can
better separate closely spaced multipath components and reduce the
estimation error. This allows the SBL-based estimates to approach the
multipath CRLB more closely. The FIM and Schur-complement derivations for the approximate reference bound are provided in Appendix~\ref{app:crlb_proof}.

\section{SBL-Assisted Robust Navigation}
\label{sec:nav}

The proposed hybrid GLRT and off-grid SBL multipath refinement provides
accurate LOS delay--Doppler estimates. These navigation parameters are tracked and refined over time via standard
tracking approaches, including delay-locked loops (DLLs),
phase-locked loops (PLLs), and Kalman filters (KFs)~\cite{kassas2024truth}. In this
framework, a KF is used to track the SBL-derived delay--Doppler
estimates~\cite{shahcheraghi2025acq}. The tracked measurements are then
used in the navigation EKF.

\subsection{Measurement and State Model}
\label{subsec:measurement_model}

The following measurement models relate the receiver state to the
states of the visible LEO satellites.

\subsubsection{Pseudorange Model}

The delay estimated by the SBL-based processing is obtained from a periodic signal structure and is therefore known only modulo the repetition period $\hat{T}_{0,i}$. The periodic-delay ambiguity is resolved using the GLRT delay estimate to determine the appropriate repetition index and obtain absolute LOS delay. The SBL estimated absolute LOS delay is converted to pseudorange as
\begin{equation}
\begin{aligned}
\rho_i(k)
=
c\,\hat{\tau}_i(k)={}&
\left\|
\mathbf{r}_r(k)
-
\mathbf{r}_{\mathrm{leo},i}
\left(t_i^{\mathrm{tx}}(k)\right)
\right\|_2
\\
&+
c\left[
\delta t_r(k)
-
\delta t_{\mathrm{leo},i}
\left(t_i^{\mathrm{tx}}(k)\right)
\right]
\\
&+
c\,\delta t_{\mathrm{iono},i}(k)
+
c\,\delta t_{\mathrm{tropo},i}(k)
+
v_{\rho_i}(k),
\end{aligned}
\label{eq:pseudorange}
\end{equation}
where $t_k=kT+t_0$ denotes the $k$th measurement epoch, with $T$ being
the measurement interval and $t_0$ the initial time. The transmission
time of the $i$th LEO signal is
$t_i^{\mathrm{tx}}(k)=t_k-\delta t_{\mathrm{TOF},i}(k)$, where
$\delta t_{\mathrm{TOF},i}(k)$ denotes the signal time-of-flight.
Moreover, $\mathbf{r}_r(k)$ and
$\mathbf{r}_{\mathrm{leo},i}(t_i^{\mathrm{tx}}(k))$ are 3-D
positions of the receiver and $i$th LEO satellite, respectively. $\delta t_r(k)$ and
$\delta t_{\mathrm{leo},i}(t_i^{\mathrm{tx}}(k))$ represent the receiver
and satellite clock biases, while
$\delta t_{\mathrm{iono},i}(k)$ and
$\delta t_{\mathrm{tropo},i}(k)$ account for ionospheric and
tropospheric delays, respectively. $v_{\rho_i}(k)$ represents
the remaining pseudorange measurement error~\cite{kassas2024adastra}.

\subsubsection{Doppler Measurement Model}

The Doppler estimated by SBL is converted to a pseudorange-rate measurement
\begin{equation}
\begin{aligned}
\dot{\rho}_i(k)
&=
-\frac{c}{f_{c,i}}\hat{\nu}_i(k)
\\
&=
\frac{
\left[
\dot{\mathbf{r}}_r(k)
-
\dot{\mathbf{r}}_{\mathrm{leo},i}
\left(t_i^{\mathrm{tx}}(k)\right)
\right]^T
\left[
\mathbf{r}_r(k)
-
\mathbf{r}_{\mathrm{leo},i}
\left(t_i^{\mathrm{tx}}(k)\right)
\right]
}{
\left\|
\mathbf{r}_r(k)
-
\mathbf{r}_{\mathrm{leo},i}
\left(t_i^{\mathrm{tx}}(k)\right)
\right\|_2
}
\\
&\quad
+c\left[
\dot{\delta t}_r(k)
-
\dot{\delta t}_{\mathrm{leo},i}
\left(t_i^{\mathrm{tx}}(k)\right)
\right]
\\
&\quad
+c\,\dot{\delta t}_{\mathrm{iono},i}(k)
+c\,\dot{\delta t}_{\mathrm{tropo},i}(k)
+v_{\dot{\rho}_i}(k).
\end{aligned}
\label{eq:pseudorange_rate}
\end{equation}
where $\dot{\mathbf{r}}_r(k)$ and
$\dot{\mathbf{r}}_{\mathrm{leo},i}(t_i^{\mathrm{tx}}(k))$ are
receiver and satellite velocity vectors, respectively.
$\dot{\delta t}_r(k)$ and
$\dot{\delta t}_{\mathrm{leo},i}(t_i^{\mathrm{tx}}(k))$ represent the
corresponding receiver and satellite clock drifts, while
$\dot{\delta t}_{\mathrm{iono},i}(k)$ and
$\dot{\delta t}_{\mathrm{tropo},i}(k)$ denote the rates of the
ionospheric and tropospheric delays.
$v_{\dot{\rho}_i}(k)$ represents the pseudorange-rate measurement error~\cite{kassas2024adastra}.

The satellite positions and velocities are obtained from their TLEs using the SGP4 propagator~\cite{shahcheraghi2025acq,khairallah2024ephemeris}, while the satellite clock bias and drift are assumed to be known. The EKF state includes receiver position, velocity, clock bias, and clock drift.\begin{equation}
\mathbf{x}=
\left[
\mathbf{r}_r^{\mathrm T},
\dot{\mathbf{r}}_r^{\mathrm T},
c\delta t_r,
c\delta\dot{t}_r
\right]^{\mathrm T}.
\label{eq:state_vector}
\end{equation}

 \subsection{EKF Prediction and Sequential Update}
\label{subsec:ekf_prediction_update}

The EKF prediction step uses a constant-velocity model for the receiver and a bias-and-drift model for the receiver clock, with unmodeled acceleration included as process noise. The state estimate and covariance are propagated as follows:
\begin{align}
\hat{\mathbf{x}}(k|k-1)
&=
\mathbf{F}(k)
\hat{\mathbf{x}}(k-1|k-1), \\
\mathbf{P}(k|k-1)
&=
\mathbf{F}(k)
\mathbf{P}(k-1|k-1)
\mathbf{F}^{\mathrm T}(k)
+
\mathbf{Q}(k),
\label{eq:prediction}
\end{align}
where $\mathbf{P}$ is state-error covariance matrix and
$\mathbf{Q}(k)$ is process noise covariance.
The
state-transition matrix is
\begin{equation}
\mathbf{F}(k)=
\begin{bmatrix}
\mathbf{I}_3 & T_k\mathbf{I}_3 & \mathbf{0}_{3\times2} \\
\mathbf{0}_{3\times3} & \mathbf{I}_3 & \mathbf{0}_{3\times2} \\
\mathbf{0}_{2\times3} & \mathbf{0}_{2\times3} &
\mathbf{F}_{\mathrm{clk}}(k)
\end{bmatrix},
\quad
\mathbf{F}_{\mathrm{clk}}(k)=
\begin{bmatrix}
1 & T_k \\
0 & 1
\end{bmatrix},
\label{eq:F_matrix}
\end{equation}
where $T_k$ is the navigation update interval. The kinematic
component of the process covariance is modeled as
\begin{equation}
\mathbf{Q}_{\mathrm{kin}}(k)
=
\sigma_a^2
\begin{bmatrix}
\frac{T_k^4}{4}\mathbf{I}_3 &
\frac{T_k^3}{2}\mathbf{I}_3 \\
\frac{T_k^3}{2}\mathbf{I}_3 &
T_k^2\mathbf{I}_3
\end{bmatrix},
\label{eq:Q_kin}
\end{equation}
where $\sigma_a$ is the standard deviation of the unmodeled
receiver acceleration. The complete process covariance is 
\begin{equation}
\mathbf{Q}(k)=
\begin{bmatrix}
\mathbf{Q}_{\mathrm{kin}}(k) & \mathbf{0}_{6\times2} \\
\mathbf{0}_{2\times6} & \mathbf{Q}_{\mathrm{clk}}(k)
\end{bmatrix},
\label{eq:Q_full}
\end{equation}
where $\mathbf{Q}_{\mathrm{clk}}(k)$ accounts for stochastic variations
in the receiver clock bias and drift. After the prediction step, measurements from visible LEO satellites are incorporated sequentially into the EKF update. The contribution of each measurement is adaptively weighted using a reliability indicator derived from the noise precision learned by SBL stage. Since the precision is estimated for each sample, its average over the CPI provides a single reliability measure for $i$th LEO signal:
\begin{equation}
\bar{\lambda}_i^{\mathrm{SBL}}(k)
=
\frac{1}{N}
\sum_{n=1}^{N}
\lambda_{i,n}^{(\infty)}(k),
\label{eq:lambda_sbl}
\end{equation}
where $\lambda_{i,n}^{(\infty)}$ is the converged noise precision of the
$n$-th observation. A larger $\bar{\lambda}_i^{\mathrm{SBL}}(k)$ is
interpreted as higher estimated reliability and is used to adapt the
measurement variance in the EKF. This quantity serves as a reliability
indicator rather than a direct estimate of the posterior uncertainty of
the delay--Doppler estimates. 

For each measurement type
$m \in \{\rho,\nu\}$, variance is
\begin{equation}
R_i^{(m)}(k)
=
\frac{
\sigma_{\mathrm{base},i}^{(m),2}
}{
\bar{\lambda}_i^{\mathrm{SBL}}(k)
},
\label{eq:sbl_weight}
\end{equation}
where $\sigma_{\mathrm{base},i}^{(m),2}$ denotes the nominal variance
of the measurement. Thus, lower SBL precision results in a larger
measurement variance and lower weight in the EKF.

The innovation is obtained from difference between the
SBL-derived measurement and its predicted value:
\begin{equation}
\zeta_i^{(m)}(k)
=
z_i^{(m)}(k)
-
h_i^{(m)}
\left(
\hat{\mathbf{x}}(k|k-1)
\right),
\label{eq:general_innovation}
\end{equation}
where $h_i^{(m)}(\cdot)$ is the corresponding measurement model introduced
in Section~\ref{subsec:measurement_model}. The measurement Jacobian is
evaluated at the predicted state, and the Kalman gain is
\begin{equation}
\begin{aligned}
\mathbf{K}_i^{(m)}(k)
={}&
\mathbf{P}(k|k-1)
\mathbf{H}_i^{(m),\mathrm T}(k)
\\
&\times
\left[
\mathbf{H}_i^{(m)}(k)
\mathbf{P}(k|k-1)
\mathbf{H}_i^{(m),\mathrm T}(k)
+
R_i^{(m)}(k)
\right]^{-1}.
\end{aligned}
\label{eq:kalman_gain_general}
\end{equation}
The state and covariance are then updated as
\begin{align}
\hat{\mathbf{x}}(k|k)
&=
\hat{\mathbf{x}}(k|k-1)
+
\mathbf{K}_i^{(m)}(k)\zeta_i^{(m)}(k),\\
\mathbf{P}(k|k)
&=
\left[
\mathbf{I}
-
\mathbf{K}_i^{(m)}(k)
\mathbf{H}_i^{(m)}(k)
\right]
\mathbf{P}(k|k-1).
\label{eq:general_ekf_update}
\end{align}
 
\section{Simulation Results}
\label{sec:results}

\subsection{Simulation Setup and Scenario Description }

The proposed framework is evaluated in a representative urban-canyon scenario using Orbcomm, Iridium NEXT, and Starlink LEO constellations. The receiver follows a vehicle motion model, incorporating constant-velocity segments and intermittent stops to mimic typical urban driving behavior. The urban channel is modeled as a Rician channel with a
dominant LOS and multiple delayed NLOS, with
Rayleigh-distributed scattered components and a Rician $K$-factor. The channel parameters are chosen to represent representative urban propagation conditions, following the modeling guidelines of 3rd generation partnership project (3GPP) TR 38.811 for non-terrestrial networks (NTN) and TR 38.901 for terrestrial urban environments~\cite{3gpp38811,3gpp38901}.
 The NLOS components are generated using a statistical sparse multipath model, with excess delays and complex gains selected within the specified ranges. This formulation provides a controlled representation of unresolved urban multipath and allows the robustness of the proposed delay--Doppler estimation method to be evaluated under diverse multipath conditions. 
 
 The received signal is also affected by temporally correlated and heavy-tailed disturbances, modeled as an AR(1) process with Student's-$t$ innovations. The key channel and noise parameters are listed in
Table~\ref{tab:simulation_environment}. In the simulation, the true propagation delay includes both geometric and atmospheric components, modeled using the Klobuchar and Saastamoinen models~\cite{klobuchar1987,saastamoinen1972}. SBL estimates the total delay without separating these effects, so residual atmospheric errors contribute to the measurement uncertainty. EKF uses the SBL-derived $\bar{\lambda}_i$ to adaptively scale the measurement covariance, reducing the impact of less reliable measurements.
 
 The simulated received LEO signals are sampled at \(f_s=10\) MHz and processed using the proposed hybrid framework, with constellation-specific carrier frequencies, bandwidths, and CPI durations for controlled delay--Doppler evaluation. Within this framework, the GLRT-based
acquisition provides initial delay--Doppler estimates, which are then
refined by the robust off-grid EM-SBL algorithm within a localized search
region. The resulting delay--Doppler estimates are subsequently
tracked and provided to the navigation EKF. The main processing
parameters for GLRT acquisition and SBL refinement are summarized in
Table~\ref{tab:processing_parameters}.
\begin{table}[!t]
\centering
\caption{LEO--SoOP signal and simulation parameters.}
\label{tab:simulation_environment}
\renewcommand{\arraystretch}{.7}
\setlength{\tabcolsep}{2pt}
\resizebox{\columnwidth}{!}{%
\begin{tabular}{lccc}
\toprule
\textbf{Parameter} & \textbf{Iridium NEXT} & \textbf{Starlink} & \textbf{Orbcomm} \\
\midrule
Carrier frequency
& $1.626$ GHz
& $11.7$ GHz
& $137.5$ MHz \\

CPI duration
& $102.4\,\mu$s
& $40\,\mu$s
& $102.4\,\mu$s \\

Nominal bandwidth
& $25$ kHz
& $5$ MHz
& $25$ kHz \\

\midrule
Rician factor ($K$)
& \multicolumn{3}{c}{$2$--$8$ dB, elevation dependent} \\

NLOS components
& \multicolumn{3}{c}{$4$--$7$, elevation dependent} \\

Multipath excess delay
& \multicolumn{3}{c}{$[50,600]$ ns} \\

\midrule
AR(1) correlation coefficient ($\rho$)
& \multicolumn{3}{c}{$0.65$} \\

Student's $t$ degrees of freedom ($\nu$)
& \multicolumn{3}{c}{$2.5$} \\

Number of satellites (visible)
& \multicolumn{3}{c}{$4$--$6$} \\

Elevation mask
& \multicolumn{3}{c}{$20^\circ$} \\
\bottomrule
\end{tabular}%
}
\end{table}
\begin{table}[!t]
\centering
\caption{Signal processing parameters for GLRT and SBL refinement.}
\label{tab:processing_parameters}
\renewcommand{\arraystretch}{.67}
\setlength{\tabcolsep}{4pt}
\begin{tabular}{l|ll}
\hline
\textbf{Stage} & \textbf{Parameter} & \textbf{Value} \\
\hline
\multirow{4}{*}{GLRT}
& Delay search range ($\Delta \tau_{\mathrm{max}})$ & $\pm 5~\mu\mathrm{s}$ \\
& Doppler search range ($\Delta \nu_{\mathrm{max}})$ & $\pm 10$ kHz \\
& Delay grid points ($M_\tau,_{\mathrm{GLRT}})$ & $101$ \\
& Doppler grid points ($M_\nu,_{\mathrm{GLRT}})$ & $201$ \\
\hline
\multirow{4}{*}{SBL}
& Fine delay search window ($\Delta \tau_{\mathrm{fine}})$ & $[-400,1000]$ ns \\
& Fine Doppler search window ($\Delta \nu_{\mathrm{fine}})$ & $\pm250$ Hz \\
& Delay grid points ($M_{\tau,\mathrm{SBL}})$ & $31$ \\
& Doppler grid points ($M_{\nu,\mathrm{SBL}})$ & $27$ \\
\hline
\multirow{3}{*}{EKF}
& Pseudorange base uncertainty ($\sigma_\rho)$ & $3.0$ m \\
& Range-rate base uncertainty ($\sigma_{\dot{\rho}})$ & $0.1$ m/s \\
& Covariance floor ($\sigma_{\mathrm{floor}}^2)$ & $50^2$ m$^2$
\\
\hline
\end{tabular}
\end{table}
The simulation compares several processing approaches under identical conditions, including
conventional acquisition, sparse recovery, and SBL-based refinement. OMP is used as a sparse-recovery baseline, while two SBL-based are considered: FMLM approach and EM-based refinement. EM-OMP and EM-FMLM are used to evaluate the additional benefit of EM-SBL refinement when initialized from OMP and FMLM, respectively. Four scenarios are considered: (i) AWGN without multipath, (ii) AWGN with multipath, (iii) AR(1) Student-$t$ noise with multipath using the classic GLRT, and (iv) AR(1) Student's-$t$ noise with multipath using robust GLRT, which reduces temporal correlation by whitening the received signal and delay--Doppler atoms with the inverse AR(1) filter. The resulting residuals are then reweighted according to the Student's-$t$ model, with large residuals assigned lower weights to reduce the effect of impulsive disturbances. The noise scale is estimated using the median absolute deviation (MAD), and the robust statistic is maximized to obtain the final estimates.
\begin{table}[!t]
\centering
\caption{GLRT acquisition results (selected SNR values).}
\label{tab:glrt_performance}
\renewcommand{\arraystretch}{1.5}
\setlength{\tabcolsep}{4pt}
\resizebox{\columnwidth}{!}{%
\begin{tabular}{c| l c c c c c c}
\toprule
\textbf{\large Scen.} 
& \textbf{\large Noise Model} 
& \textbf{\large Multipath} 
& \textbf{\large GLRT Type} 
& \textbf{\large SNR (dB)} 
& \textbf{\large \(P_D\)}  
& \textbf{\large RMSE\(_\tau\) (ns)} 
& \textbf{\large RMSE\(_\nu\) (Hz)} \\
\midrule

\multirow{3}{*}{\large S1} 
& \multirow{3}{*}{\large AWGN} 
& \multirow{3}{*}{\large No} 
& \multirow{3}{*}{\large Classic} 
& {\large -10} & {\large 0.97} & {\large 14.7}  & {\large 2390.8} \\
& & & & {\large 0}   & {\large 1.00} & {\large 7.7}   & {\large 727.0}  \\
& & & & {\large +10} & {\large 1.00} & {\large 5.7}   & {\large 307.6}  \\

\midrule

\multirow{3}{*}{\large S2} 
& \multirow{3}{*}{\large AWGN} 
& \multirow{3}{*}{\large Yes} 
& \multirow{3}{*}{\large Classic} 
& {\large -10} & {\large 0.95} & {\large 139.4} & {\large 1952.1} \\
& & & & {\large 0}  & {\large 1.00} & {\large 182.1} & {\large 940.2}  \\
& & & & {\large +10} & {\large 1.00} & {\large 163.4} & {\large 711.7}  \\

\midrule

\multirow{3}{*}{\large S3} 
& \multirow{3}{*}{\large AR(1) \; Student's-$t$} 
& \multirow{3}{*}{\large Yes} 
& \multirow{3}{*}{\large Classic} 
& {\large -10} & {\large 0.85} & {\large 447.2} & {\large 2945.2} \\
& & & & {\large 0}   & {\large 1.00} & {\large 172.6} & {\large 1062.0} \\
& & & & {\large +10} & {\large 1.00} & {\large 172.9} & {\large 701.3}  \\

\midrule

\multirow{3}{*}{\large S4} 
& \multirow{3}{*}{\large AR(1) \; Student's-$t$} 
& \multirow{3}{*}{\large Yes} 
& \multirow{3}{*}{\large Robust} 
& {\large -10} & {\large 1.00} & {\large 175.6} & {\large 1380.9} \\
& & & & {\large 0}   & {\large 1.00} & {\large 142.8} & {\large 759.5}  \\
& & & & {\large +10} & {\large 1.00} & {\large 176.2} & {\large 710.9}  \\

\bottomrule
\end{tabular}%
}
\end{table}
\begin{figure}
            \centering
            \includegraphics[width=.74\linewidth]{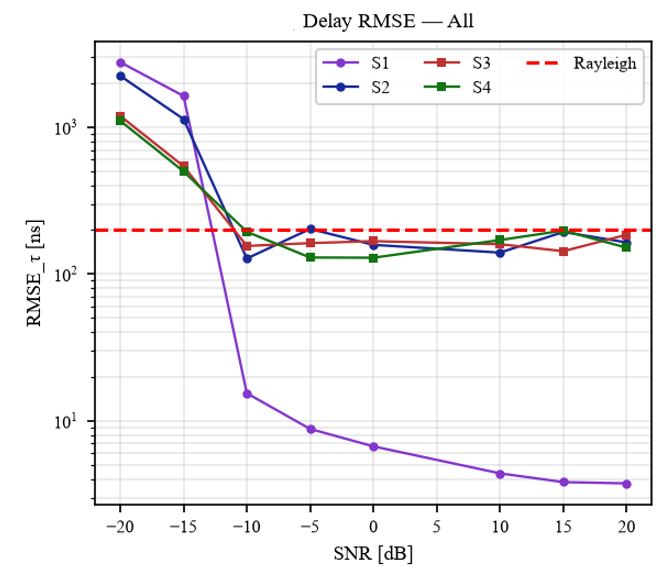}
            \caption{Delay RMSE of the GLRT-based acquisition versus SNR.}
            \label{fig:delay}
        \end{figure}
\subsection{Acquisition and Tracking Performance}
\label{subsec:acq_tracking}

The conventional GLRT-based acquisition results for the considered channel conditions are summarized in Table~\ref{tab:glrt_performance}. The detection threshold corresponds to $P_{\mathrm{FA}}=10^{-4}$~\cite{kay1998detection}, and $P_D$ denotes the fraction of correctly detected trials.
In the multipath-free scenarios, the GLRT performs as expected, with reliable detection (\(P_D\)=1) and decreasing delay RMSE as the SNR increases. For the multipath scenarios, however, the delay RMSE remains substantially higher across the considered SNR range, despite reliable detection. This behavior indicates a systematic estimation bias associated with the single-peak acquisition model. The observed behavior is consistent with the theoretical analysis in Section~\ref{sec:glrt_analysis}, the CAF-based GLRT selects the strongest correlation peak, which may be displaced from the true LOS delay by unresolved NLOS components. Consequently, increasing the SNR does not fully eliminate the resulting delay error, and the RMSE tends to approach a persistent error floor, as observed in Fig.~\ref{fig:delay}. These results highlight the limitation of GLRT-based acquisition for accurate LOS estimation in multipath environments and motivate the evaluation of the SBL-based refinement.

\begin{figure}[!t]
    \centering
\includegraphics[width=\columnwidth]{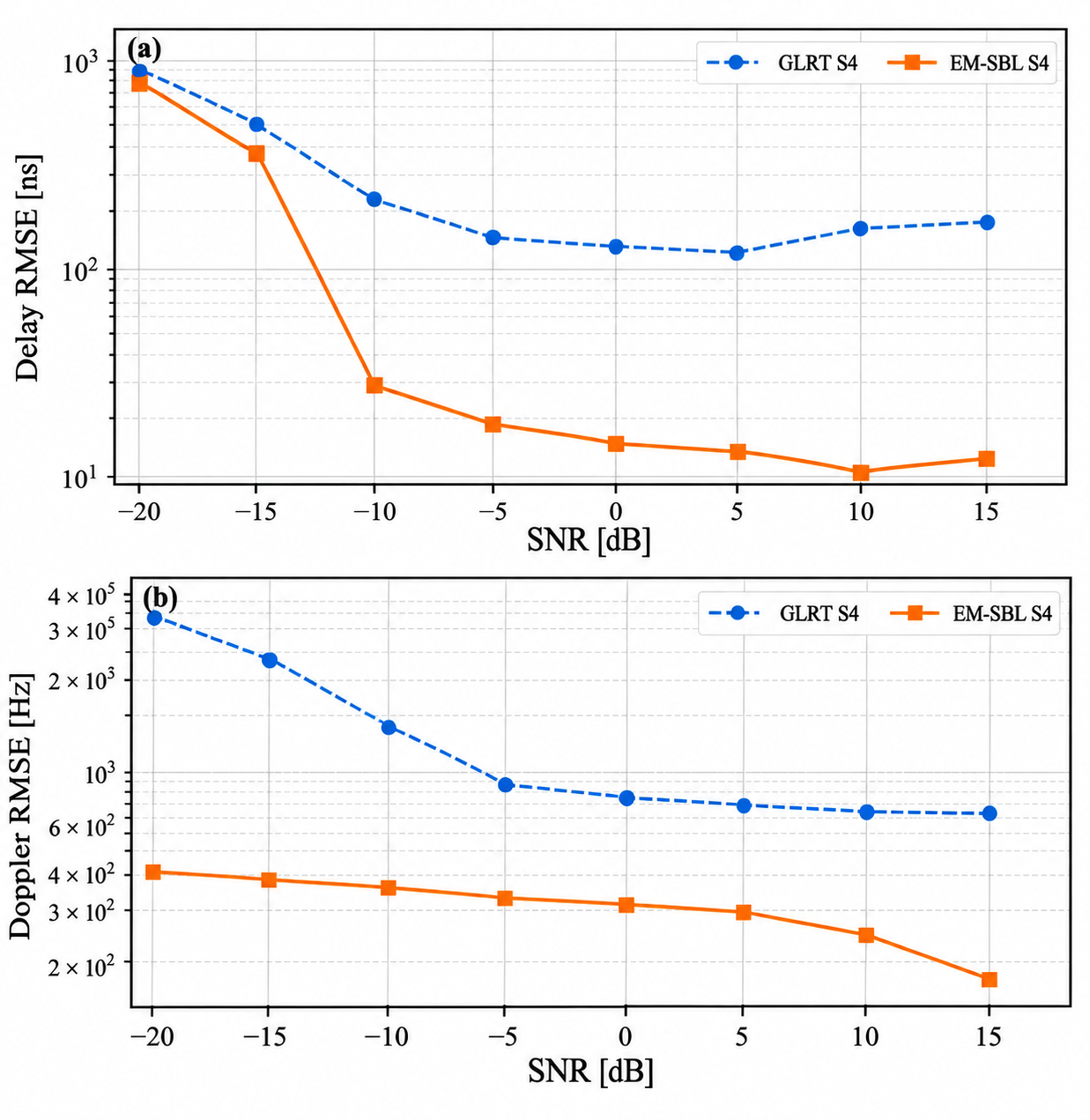}

    \caption{Delay--Doppler estimation performance of GLRT and EM-SBL in Scenario~(iv): (a) delay RMSE and (b) Doppler RMSE.}
    \label{fig:Capture122}
\end{figure}

 The performance of the proposed off-grid SBL framework is summarized in Fig.~\ref{fig:Capture122} and
Table~\ref{tab:comprehensive_fixed} for Scenario~(iv). Compared with GLRT,
OMP, and FMLM, the SBL-based methods achieve lower delay and Doppler RMSEs
across the considered SNR range. The EM-SBL variants generally provide substantial improvements over GLRT, by
combining ARD with EM-based refinement to identify the dominant paths and
refine their delay--Doppler parameters beyond the initial grid. EM-SBL reduces the delay RMSE by up to 80\%, providing a more accurate LOS estimate, which leads to more accurate measurements for navigation filter.

\begin{table*}[!t]
\centering
\footnotesize
\caption{Comprehensive performance comparison of different methods in Scenario~(iv).}
\label{tab:comprehensive_fixed}
\renewcommand{\arraystretch}{0.95}
\setlength{\tabcolsep}{3pt}
\resizebox{\textwidth}{!}{%
\begin{tabular}{ll|ccccccccc}
\toprule
\textbf{Metric} & \textbf{Parameters}
& \textbf{GLRT}
& \textbf{OMP}
& \textbf{FMLM}

& \textbf{EM-SBL (off-grid,OMP)}
& \textbf{EM-SBL (on-grid,FMLM)}
& \textbf{EM-SBL (off-grid,FMLM)} \\
\midrule

\multirow{3}{*}{\textbf{Delay RMSE [ns] (B=5MHz,K=8)}}
& SNR = $-10$ dB
& 175.6 & 44.0 & 43.6 & 42.3 & 42.1   & 41.7 \\

& SNR = $0$ dB
& 142.8 & 39.6 & 45.8  & 35.9 & 35.4 & 34.3 \\

& SNR = $+10$ dB
& 176.2 & 49.0 & 50.1 & 34.1 & 34.9   & 32.9 \\

\midrule

\textbf{Improvement vs GLRT (SNR=10,K=8,B=5MHz)}
&
& -- & +65.6\% & +65\% & +76.1  & +75.5\% & +77\% \\

\midrule

\multirow{3}{*}{\textbf{Doppler RMSE [Hz] (B=5MHz,K=8)}}
& SNR = $-10$ dB
& 1380.9 & 354.2 & 339.0  & 313.2 & 329.0 & 317.6 \\

& SNR = $0$ dB
 & 759.5 & 321.7 & 293.8 & 274.0 & 249.9 & 247.8 \\

& SNR = $+10$ dB
& 710.9 & 241.5 & 287.2  & 271.7 & 277.0 & 259.2 \\

\midrule

\multirow{3}{*}{\textbf{Bandwidth Sensitivity [ns] (SNR=0,K=8)}}
& $B = 1$ MHz
& 318.2 & 117.5 & 109.4 & 93.8 & 98.6 & 95.0 \\

& $B = 10$ MHz
& 157.6 & 39.1 & 41.2 & 34.7 & 35.2 & 33.9 \\

& $B = 20$ MHz
& 100.8 & 34.9  & 37.1 & 31.3 & 34.4 & 33.7 \\

\midrule

\multirow{3}{*}{\textbf{$K$-factor bias [ns] (SNR=0,B=5MHz)}}
& $K = 0$ dB
& 378.0 & 109.7 & 112.4 & 106.3 &  108.0 & 108.0 \\

& $K = 2$ dB
& 256.5 & 85.3 & 84.5 & 79.1 &  80.9 & 80.1 \\
& $K = 10$ dB
& 126.5 & 35.4 & 49.2 & 34.8 &  35.3 & 34.5 \\
\midrule

\textbf{Runtime [ms]}
&
& 0.29 & 1.25 & 3.40 & 8.5  & 8.4 & 8.7 \\
\midrule
\textbf{Computational Complexity}
&
& O(NG) & O(KNG) & $O\left(I(K^2G+NK)\right)$ & $O\left(I(K^2G+NK)\right) + O(G)$  & $O\left(I(K^2G+NK)\right)$ & $O\left(I(K^2G+NK)\right) + O(G)$ \\
\bottomrule
\end{tabular}%
}
\end{table*}

\subsection{SBL-Aided Navigation Performance}
\label{subsec:navigation_results}
\begin{figure}
    \centering
    \includegraphics[width=.95\linewidth]{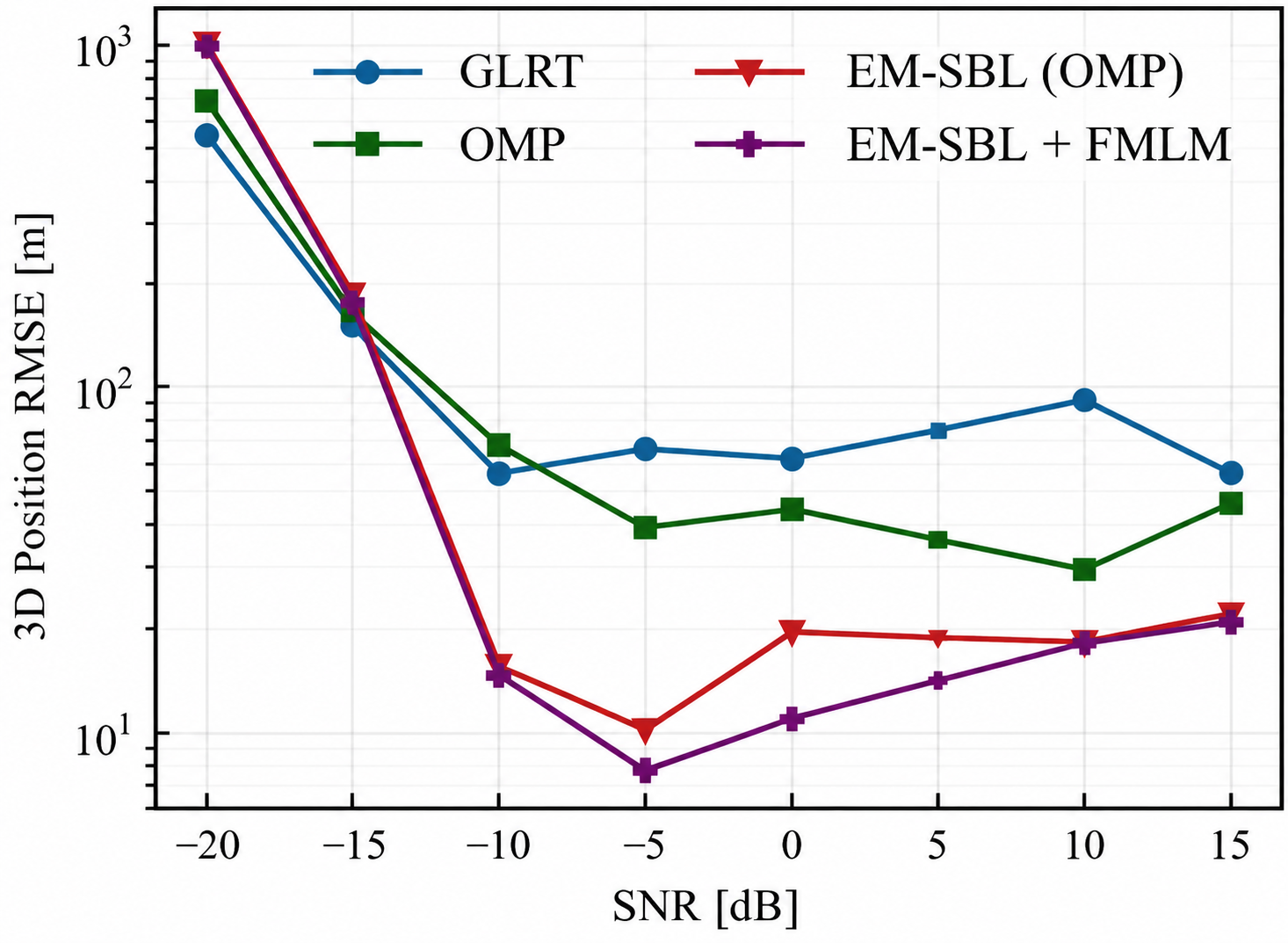}
    \caption{ 3D position RMSE versus SNR for different methods in Scenario~(iv).}
    \label{fig:P3b_Position_RMSE_paper}
\end{figure}

The navigation results show how delay--Doppler estimation accuracy affects
positioning performance. As shown in Table~\ref{tab:navigation_performance}, SBL-based estimators achieve lower position RMSE than the GLRT baseline
across the considered SNR range. At \(0\) dB, EM--SBL reduces the position
RMSE from \(76.2\)m with GLRT to \(10.9\)m. At \(+10\) dB, EM--SBL
achieves the lowest RMSE of \(18.2\) m, compared with \(92.6\) m for GLRT,
corresponding to a reduction of approximately \(88.5\%\). The 3D position RMSE in Fig.~\ref{fig:P3b_Position_RMSE_paper} also shows lower errors for the SBL-based methods than for GLRT across the considered SNR range. These results demonstrate the overall benefit of the proposed SBL-based processing under the urban multipath and non-Gaussian noise conditions.

\begin{table}[t]
\centering
\caption{Navigation performance in Scenario~(iv).}
\label{tab:navigation_performance}
\renewcommand{\arraystretch}{1.15}
\setlength{\tabcolsep}{3pt}
\resizebox{\columnwidth}{!}{%
\begin{tabular}{c|rrr|rr}
\toprule
\multirow{2}{*}{\textbf{SNR (dB)}} 
& \multicolumn{3}{c|}{\textbf{Position RMSE (m)}} 
& \multicolumn{2}{c}{\textbf{Improvement (\%)}} \\
\cline{2-6}
& \textbf{GLRT} 
& \textbf{EM--OMP} 
& \textbf{EM--FMLM} 
& \textbf{EM--OMP} 
& \textbf{EM--FMLM} \\
\midrule
\(-10\) & 56.4 & 15.6 & 14.6 & 72.4 & 74.2 \\
\(-5\) & 66.4 & 10.2 & 7.6 & 84.7 & 88.5 \\
\(0\)   & 76.2 & 19.5 & 10.9 & 68.7 & 82.5 \\
\(+5\) & 56.4 & 18.5 & 15.5 & 67.2 & 73.7 \\
\(+10\) & 92.6 & 18.2 & 18.4 & 80.3 & 80.2 \\
\bottomrule
\end{tabular}%
}
\end{table}

\section{Conclusion}
\label{sec:conclusion}

This paper presented a hybrid GLRT and off-grid EM-SBL framework for multipath-aware navigation in urban environments. The framework combines GLRT-based acquisition with EM-SBL refinement to separate LOS from multipath and refine delay--Doppler parameters. The robust noise model improves estimation under heavy-tailed and temporally correlated disturbances, while the off-grid formulation reduces discretization errors. Under dense multipath, the proposed SBL-based estimators substantially reduce LOS delay and Doppler errors compared with GLRT, enabling up to \(88.5\%\) lower navigation errors when integrated into the EKF with SBL-derived reliability information.

\section*{Acknowledgment}

The authors used OpenAI ChatGPT for language editing and clarity improvement. All content was reviewed and verified by the authors, who remain responsible for the final manuscript.

\appendices

\section{ Analysis of GLRT Delay Bias Under Multipath}
\label{app:bias_proof}

Under the multipath model in \eqref{eq:received_signal}, consider the GLRT delay
statistic for a fixed Doppler estimate $\hat{\nu}$:
\begin{equation}
\Lambda(\tau)
=
\frac{
\left|
\mathbf{d}^{H}(\tau,\hat{\nu})\mathbf{y}
\right|^2
}{
\|\mathbf{d}(\tau,\hat{\nu})\|_2^2
}.
\label{eq:app_glrt_stat}
\end{equation}
For local delay analysis, $\hat{\nu}$ is assumed close to the true Doppler. Since the noise variance and dictionary norm are locally delay-independent, they do not affect the maximum location. The normalized CAF is defined as
\begin{equation}
C(\tau;\tau_p,\nu_p)
=
\frac{
\mathbf{d}^{H}(\tau,\hat{\nu})
\mathbf{d}(\tau_p,\nu_p)
}{N}.
\label{eq:app_caf}
\end{equation}

Substituting the multipath model into \eqref{eq:app_glrt_stat} and
taking the expectation with respect to the noise $\mathbb{E}_{n}[\cdot]$ gives
\begin{equation}
\mathbb{E}_{n}[\Lambda(\tau)]
=
\left|
\sum_{p=0}^{P}
a_p C(\tau;\tau_p,\nu_p)
\right|^2
+
\frac{\sigma^2}{N},
\label{eq:app_expected_glrt}
\end{equation}
 Since the last term is independent of $\tau$, it does not affect
the location of the maximum. Let $\tau_0$ is the LOS delay and
$\Delta\tau_p=\tau_p-\tau_0$ is the excess delay of the $p$th
multipath. For unresolved multipath, we assume
$|\Delta\tau_p|B<1.
\label{eq:app_unresolved}$
Thus, all unresolved paths
lie within a local delay region around the LOS. In this region, the CAF is assumed smooth and approximately quadratic around its maximum. The CAF of each multipath can therefore be expanded around $\tau_0$ as
\begin{equation}
C(\tau;\tau_p,\nu_p)
\approx
C(\tau;\tau_0,\nu_p)
+
\Delta\tau_p
\left.
\frac{\partial C(\tau;\tau',\nu_p)}
{\partial\tau'}
\right|_{\tau'=\tau_0}.
\label{eq:app_caf_expansion}
\end{equation}

To determine the shift of the maximum, define $f(\tau)=\mathbb{E}_{n}[\Lambda(\tau)]$ and use Taylor expansion around $\tau_0$:
\begin{equation}
f(\tau)
\approx
f(\tau_0)
+
f'(\tau_0)(\tau-\tau_0)
+
\frac{1}{2}f''(\tau_0)(\tau-\tau_0)^2.
\label{eq:app_taylor}
\end{equation}

Expanding \eqref{eq:app_expected_glrt} gives individual-path and cross terms. For closely spaced unresolved paths, the cross terms are assumed negligible near the dominant peak. Thus, the first and second derivatives are approximated using the individual-path contributions. Assuming a locally symmetric CAF with negligible Doppler mismatch and nearly constant curvature over the region in \eqref{eq:app_unresolved}, define
$$
K=-\left.\frac{\partial^2}{\partial\tau^2}
|C(\tau;\tau_0,\nu_0)|^2\right|_{\tau=\tau_0}>0.
$$

The first and second-order are approximated as
\begin{equation}
f'(\tau_0)
\simeq
K\sum_{p=0}^{P}|a_p|^2\Delta\tau_p,
\qquad
f''(\tau_0)
\simeq
-K\sum_{p=0}^{P}|a_p|^2.
\label{eq:app_derivatives}
\end{equation}
Here, $\Delta\tau_0=0$ for the LOS component. The factor $K$ follows from the local constant-curvature approximation and may vary along the delay axis. Setting derivative of $f(\tau)$ to zero gives
\begin{equation}
\hat{\tau}_{\mathrm{GLRT}}
\approx
\tau_0
-
\frac{f'(\tau_0)}
{f''(\tau_0)}
\approx
\frac{
\sum_{p=0}^{P}|a_p|^2\tau_p
}{
\sum_{p=0}^{P}|a_p|^2
}.
\label{eq:app_delay_estimate}
\end{equation}

Therefore, the first-order multipath-induced delay bias is
\begin{equation}
b_{\mathrm{MP}}^{\tau}
=
\hat{\tau}_{\mathrm{GLRT}}-\tau_0
\simeq
\frac{
\sum_{p=1}^{P}|a_p|^2\Delta\tau_p
}{
|a_0|^2+\sum_{p=1}^{P}|a_p|^2
}.
\label{eq:app_bias}
\end{equation}

The bias can be expressed using the Rician \(K\)-factor as
\begin{equation}
K_{\mathrm{lin}}
=
\frac{|a_0|^2}
{\sum_{p=1}^{P}|a_p|^2}.
\label{eq:app_rician}
\end{equation}
Here, $K_{\mathrm{lin}}$ is the LOS-to-NLOS power ratio. For equal-power NLOS paths with average excess delay, \eqref{eq:app_bias} becomes

\begin{equation}
\operatorname{bias}(\hat{\tau}_{\mathrm{GLRT}})
\approx
\frac{\Delta\tau}
{1+K_{\mathrm{lin}}}.
\label{eq:app_final_bias}
\end{equation}

\section{Cram\'er--Rao Lower Bound Analysis}
\label{app:crlb_proof}
The mean received signal is
\begin{equation}
\boldsymbol{\mu}_i
=
\sum_{p=1}^{P_i}
a_{i,p}\mathbf d_i(\tau_{i,p},\nu_{i,p}),
\end{equation}
where $a_{i,p}$, $\tau_{i,p}$, and $\nu_{i,p}$ denote the path amplitude,
delay, and Doppler shift, respectively. The parameters are
\begin{equation}
\boldsymbol{\vartheta}_i
=
[\tau_{i,1},\ldots,\tau_{i,P_i},
\nu_{i,1},\ldots,\nu_{i,P_i}]^{T},
\end{equation}
while the unknown path amplitudes are treated as nuisance parameters. For the complex Gaussian model with known $\boldsymbol{\Lambda}_i$, the
Slepian--Bangs formula gives~\cite{kay1993}
\begin{equation}
[\mathbf J_i]_{mn}
=
2\operatorname{Re}
\left\{
\frac{\partial\boldsymbol{\mu}_i^{H}}
{\partial\theta_m}
\boldsymbol{\Lambda}_i
\frac{\partial\boldsymbol{\mu}_i}
{\partial\theta_n}
\right\}.
\label{eq:crlb_fim}
\end{equation}
Let $\boldsymbol{\eta}_i$ contain the real and imaginary parts of the
unknown path amplitudes, and define the complete parameter vector as
\begin{equation}
\boldsymbol{\theta}_i
=
[\boldsymbol{\vartheta}_i^{T},\boldsymbol{\eta}_i^{T}]^{T}.
\end{equation}
Partitioning the FIM accordingly gives
\begin{equation}
\mathbf J_i
=
\begin{bmatrix}
\mathbf J_{\vartheta\vartheta,i} &
\mathbf J_{\vartheta\eta,i} \\
\mathbf J_{\eta\vartheta,i} &
\mathbf J_{\eta\eta,i}
\end{bmatrix}.
\end{equation}

For the delay--Doppler parameters, define
\begin{equation}
\mathbf D_i^\tau
=
\left[
a_{i,1}\dot{\mathbf d}_{i,1}^{\tau},
\ldots,
a_{i,P_i}\dot{\mathbf d}_{i,P_i}^{\tau}
\right],
\
\mathbf D_i^\nu
=
\left[
a_{i,1}\dot{\mathbf d}_{i,1}^{\nu},
\ldots,
a_{i,P_i}\dot{\mathbf d}_{i,P_i}^{\nu}
\right].
\end{equation}
Thus,
\begin{equation}
\mathbf J_{\vartheta\vartheta,i}
=
2\operatorname{Re}
\begin{bmatrix}
(\mathbf D_i^\tau)^H\boldsymbol{\Lambda}_i\mathbf D_i^\tau &
(\mathbf D_i^\tau)^H\boldsymbol{\Lambda}_i\mathbf D_i^\nu \\
(\mathbf D_i^\nu)^H\boldsymbol{\Lambda}_i\mathbf D_i^\tau &
(\mathbf D_i^\nu)^H\boldsymbol{\Lambda}_i\mathbf D_i^\nu
\end{bmatrix}.
\label{eq:crlb_fim_blocks}
\end{equation}

Eliminating the nuisance amplitudes through the Schur complement gives
the effective information matrix
\begin{equation}
\mathbf J_{\mathrm{eff},i}
=
\mathbf J_{\vartheta\vartheta,i}
-
\mathbf J_{\vartheta\eta,i}
\mathbf J_{\eta\eta,i}^{-1}
\mathbf J_{\eta\vartheta,i}.
\label{eq:crlb_schur}
\end{equation}
Hence, the multipath CRLB is
\begin{equation}
\operatorname{Cov}(\hat{\boldsymbol{\vartheta}}_i)
\succeq
\mathbf J_{\mathrm{eff},i}^{-1}.
\label{eq:crlb_general}
\end{equation}
Thus, the diagonal entries of $\mathbf J_{\mathrm{eff},i}^{-1}$ give the variance bounds for individual path delays and Doppler estimates, while the off-diagonal terms capture inter-path coupling. For Student's t model, the same bound can be evaluated conditionally
for a fixed precision realization $\boldsymbol{\Lambda}_i$:
\begin{equation}
\mathrm{CRLB}_{\mathrm{cond}}
(\boldsymbol{\vartheta}_i\mid\boldsymbol{\Lambda}_i)
=
\mathbf J_{\mathrm{eff},i}^{-1}.
\label{eq:crlb_conditional}
\end{equation}
This is a conditional bound, not the marginal Student's-t CRLB, since the variables are estimated within EM. Doppler CRLB propagates to pseudorange-rate
\begin{equation}
\mathrm{CRLB}
\left(\dot{\rho}_i^{\,\mathrm{meas}}\right)
=
\left(\frac{c}{f_{c,i}}\right)^2
\mathrm{CRLB}(\nu_i).
\label{eq:crlb_rhodot}
\end{equation}

\end{document}